\documentclass[preprint,12pt]{elsarticle}

\usepackage{amssymb}
\usepackage{amsmath}
\usepackage{natbib}
\usepackage{hyperref}
\usepackage{geometry}
\usepackage{array}
\usepackage{longtable}
\usepackage{graphicx}
\usepackage{booktabs}   % For \toprule, \midrule, \bottomrule
\usepackage{tcolorbox}
\usepackage{listings}
\usepackage{xcolor}
\usepackage{rotating}
\usepackage{multirow}   
\usepackage{threeparttable}
\usepackage{pifont} % 放在导言区
\usepackage{enumitem} % 导言区引入
\usepackage{caption}
\usepackage{float}  % 可选，有时辅助 float 管理

\newcommand{\cmark}{\ding{51}} % ✔
\newcommand{\xmark}{\ding{55}} % ✘

\journal{Journal of Biomedical Informatics}

\begin{document}

\begin{frontmatter}



\title{A Clinically Informed Two-Stage Framework for Renal CT Report Generation}

%% Authors
\author[ufhobi]{Renjie Liang, MSc}
\author[ufhobi]{Zhengkang Fan, MSc}
\author[ufhobi]{Jinqian Pan, MSc}
\author[ufhobi]{Chenkun Sun, MSc}
\author[ufurology]{Bruce Daniel Steinberg, MD}
\author[ufurology]{Russell Terry, MD}
\author[ufhobi]{Jie Xu, PhD\corref{cor1}}
% \cortext[cor1]{Corresponding author. Email: jie.xu@ufl.edu}

%% Affiliations
\affiliation[ufhobi]{
    organization={Department of Health Outcomes and Biomedical Informatics, University of Florida},
    addressline={2004 Mowry Road},
    city={Gainesville},
    postcode={32611},
    state={FL},
    country={USA}
}

\affiliation[ufurology]{
    organization={Department of Urology, University of Florida},
    addressline={1600 SW Archer Road},
    city={Gainesville},
    postcode={32611},
    state={FL},
    country={USA}
}

\begin{abstract}
\noindent \textbf{Objective}
Renal cancer is a common malignancy and an important cause of cancer-related mortality. Computed tomography (CT) plays a central role in early detection, staging, and treatment planning. The growing volume of CT studies increases radiologists' workload and requires specialized training to reliably interpret images and document findings. However, automatically generating such reports remains challenging as it requires integrating visual interpretation with clinical reasoning. Advances in artificial intelligence (AI), including large language models (LLMs) and vision-language systems, may help reduce radiologists’ reporting burden, enhance diagnostic accuracy, and ultimately improve patient care and clinical efficiency.

\noindent\textbf{Methods}
We propose a clinically informed, two-stage framework for automatic radiology report generation from single-slice renal CT images. 
In the first stage, we detect structured \textit{clinical features} from each 2D image using a multi-task learning model. In the second stage, we generate free-text reports conditioned on both the input image and the \textit{detected clinical features}, leveraging a vision-language foundation model. 
To assess clinical accuracy and consistency, we further extract \textit{generated clinical features} from model-generated reports and compare them with expert-annotated ground truth. We use an expert-labeled dataset for training and evaluation.

\noindent\textbf{Results}
Experiments on an expert-labeled dataset show that incorporating detected clinical features improves both the quality and clinical fidelity of the generated reports. The model achieved an average AUC of 0.75 across key imaging features and reached a maximum METEOR score of 0.33 in report generation, indicating enhanced clinical faithfulness and reduced template-driven hallucinations.

\noindent\textbf{Conclusion}
Coupling structured clinical feature detection with conditioned report generation offers a promising approach to integrate structured prediction with free-text drafting for renal CT reporting, enhancing interpretability and clinical fidelity. The results also highlight the importance of clinically meaningful evaluation metrics when developing medical AI systems.
\end{abstract}

\end{frontmatter}

\section{Introduction}
\label{sec:Introduction}

Renal cancer, also known as kidney cancer, is a malignancy with increasing global incidence. Each year, an estimated 400{,}000 new cases are diagnosed, resulting in approximately 175{,}000 deaths worldwide \cite{cirillo2024global}. Projections suggest that incidence rates will continue to rise over the next decade, underscoring the public health significance of this disease. In the United States alone, more than 81{,}000 new cases and over 14{,}000 deaths are expected in 2024 \cite{Siegel2024}. Although mortality rates have declined by 1--2\% annually due to advances in treatment, overall incidence continues to increase \cite{Siegel2024}.
The widespread adoption of cross-sectional imaging—particularly computed tomography (CT)—has led to increased incidental detection of early-stage renal tumors \cite{padala2020epidemiology}. Early diagnosis is critical, as renal cancer is often asymptomatic in its initial stages but can progress rapidly if untreated. Accurate and timely interpretation of renal CT scans is essential for clinical decision-making, including diagnosis, staging, and treatment planning. However, the growing volume of imaging studies, coupled with inter-observer variability among radiologists, poses significant challenges. These limitations highlight the need for intelligent systems capable of supporting or augmenting radiology report generation from renal CT images, which may improve diagnostic efficiency, reducing variability, and ultimately enhance patient care.

Recent advances in large model technologies, including large language models (LLMs) and vision-language models (VLMs), have transformed biomedical informatics and computer-aided diagnosis. Models such as GPT-4 \cite{achiam2023gpt}, PaLM \cite{chowdhery2023palm}, and Med-PaLM \cite{singhal2023large} have demonstrated the ability to understand clinical narratives, answer medical questions, and even provide diagnostic suggestions, functioning as virtual clinical assistants in various scenarios \cite{singhal2023large, lee2023benefits, zhou2024comprehensive}. Similarly, vision-language models (VLMs) such as BioViL~\cite{bannur2023learning} and LLaVA-Med~\cite{li2023llava} integrate visual and textual modalities to support medical image interpretation, question answering, and image-report alignment \cite{huang2025survey, liu2023visual}.
These technological breakthroughs have led to a surge of interest in medical imaging applications, particularly in automating radiology workflows. One rapidly growing area is radiology report generation, where deep models learn to produce textual reports directly from medical images. Most existing work in this space has focused on chest X-rays, leveraging large-scale public datasets such as MIMIC-CXR \cite{johnson2019mimic} and CheXpert \cite{irvin2019chexpert}.

Despite growing interest in radiology report generation, relatively little attention has been given to anatomical regions beyond the chest. In particular, renal imaging remains an underexplored area. Most existing studies on kidney CT or MRI primarily focus on classification tasks, such as tumor detection or subtype prediction, rather than comprehensive report generation \cite{alzu2022kidney, islam2022vision, uhm2021deep, yildirim2021deep}. Several challenges contribute to this gap.
First, there is a lack of large-scale, publicly available renal CT datasets with paired expert-authored reports, limiting model development and evaluation. Second, unlike chest imaging, renal CT lacks a widely accepted or standardized reporting template, resulting in a more diverse and ambiguous target output space. Third, evaluating the quality of generated radiology reports remains inherently difficult. Conventional natural language generation (NLG) metrics—such as BLEU, ROUGE, and METEOR—primarily measure lexical overlap and often fail to capture clinically meaningful content. In clinical practice, even minor discrepancies can substantially alter the diagnostic interpretation \cite{nishino2023prevalence}.

\begin{figure}[th]
    \footnotesize
    \centering
    \includegraphics[width=\textwidth]{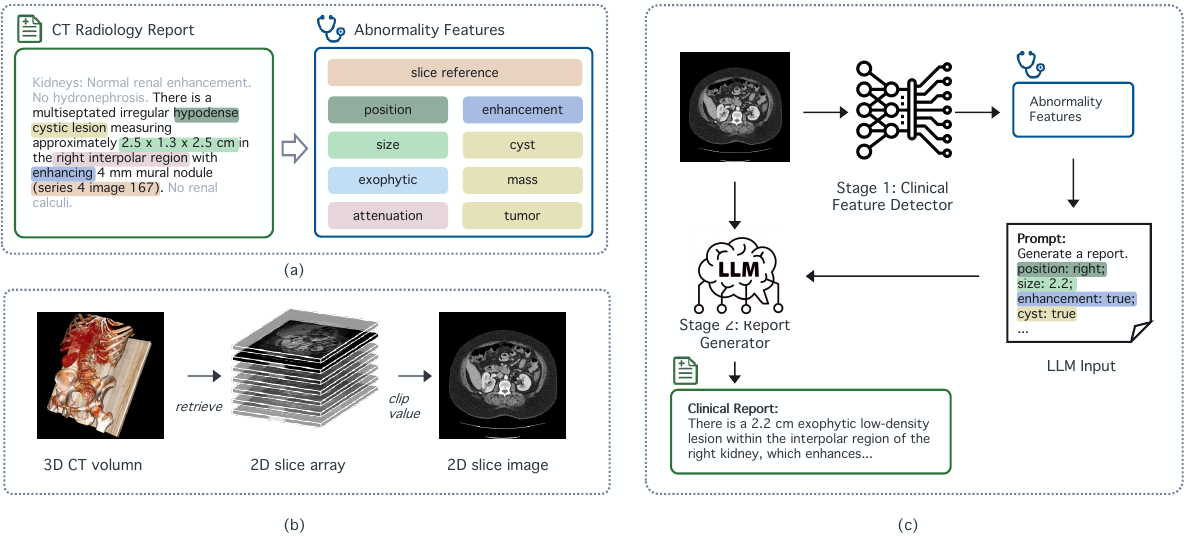}
    \caption
    {
    Dataset curation and model framework.
    (a) From expert renal CT radiology reports we derive structured\textit{ clinical (abnormality) features}—position, size, exophytic growth, attenuation, enhancement, cyst, mass, tumor—and a reference slice index; these align each 2D CT slice with labels. (b) From a 3D CT volume, we retrieve the labeled slice and apply intensity clipping to form the 2D input image. (c) Two-stage pipeline: Stage 1 detects clinical features from the image; Stage 2 conditions an LLM-based report generator on the image and the detected features via a structured prompt to produce a narrative clinical report.
    }    
    \label{fig:model_pipeline}
\end{figure}

In this study, we explore automated renal radiology report generation using real-world clinical data from the UF Health Integrated Data Repository (IDR). We construct a custom structured schema—guided by expert radiologists—for describing renal CT findings.
We introduce a two-stage pipeline: first, key radiological features—referred to as \textit{clinical features}—are detected from individual 2D CT slices; second, a vision-language model generates free-text reports conditioned on both the detected features and the corresponding CT slice. This modular design enhances interpretability and clinical relevance while enabling separate optimization and evaluation of feature extraction and narrative generation.

To our knowledge, this study represents one of the first systematic efforts in generative modeling for kidney imaging. It establishes a clinically informed framework integrating multimodal learning and domain-specific supervision, laying the groundwork for future kidney-specific AI applications in radiology and related domains.

\begin{table}[ht]
\centering
\footnotesize
\caption*{Statement of Significance}
\begin{tabular}{m{0.23\linewidth} m{0.77\linewidth}}
\toprule
\textbf{Problem} & Writing renal CT reports is time-consuming and may omit important details. Automatic solutions are needed to improve efficiency and completeness and ultimately support clinicians. \\
\hline
\textbf{Already Known} & Most radiology report generation studies have focused on chest X-rays with promising results. However, applications to renal CT remain underexplored. \\
\hline
\textbf{This Paper Adds} & We introduced the first system for renal CT report generation using a clinically informed two-stage pipeline. The method detects structured features and generates coherent reports aligned with expert findings, evaluated on real-world data from UF Health IDR. \\
\hline
\textbf{Benefit} & Radiologists and nephrologists can benefit from improved report consistency and reduced documentation burden. AI researchers may use this framework to extend multimodal generation to underexplored anatomical regions. \\
\bottomrule
\end{tabular}
\end{table}

\section{Related Works}

\subsection{Renal Abnormality Diagnosis}

Deep learning approaches have shown promise in diagnosing kidney abnormalities from CT scans. For example, Alzu'bi et al.~\cite{alzu2022kidney} proposed a method for classifying renal tumors, cysts, and stones from 2D CT slices using a newly introduced dataset. Similarly, Islam et al.~\cite{islam2022vision} developed vision transformer and transfer learning models to distinguish normal kidneys from those with tumors, cysts, or stones, also based on 2D CT images. Yildirim et al.~\cite{yildirim2021deep} focused specifically on detecting kidney stones using coronal CT slices, employing a deep learning model for automated identification. Uhm et al.~\cite{uhm2021deep} addressed a more complex task by classifying kidney cancer into five subtypes using multi-phase 3D CT scans.
While these methods demonstrate promising diagnostic performance, they primarily focus on high-level classification tasks and do not capture detailed radiological findings. None of the above studies aim to generate comprehensive radiology reports that mimic expert-written narratives. In contrast, our work detects fine-grained clinical features and generates interpretable, report-style outputs that closely mimic real-world clinical documentation.

\subsection{Large Language Models}

LLMs have shown impressive capabilities in natural language understanding and generation, with growing applications in biomedicine. OpenAI's GPT-4, for instance, performs competitively on medical licensing exams and clinical question-answering tasks \cite{nori2023capabilities}. Meta's LLaMA series, particularly LLaMA-2 and its biomedical adaptation MedLLaMA, have been fine-tuned for healthcare tasks through domain-specific pretraining and instruction tuning \cite{singhal2025toward}. Qwen, an open-source foundation model developed by Alibaba, has introduced medical variants that contribute to the multilingual expansion of biomedical LLMs \cite{bai2023qwen}.
Beyond general-purpose LLMs, several models have been explicitly designed for biomedical use. Examples include BioGPT \cite{luo2022biogpt}, trained on PubMed abstracts, and PMC-LLaMA \cite{wu2024pmc}, which integrates biomedical knowledge into the LLaMA architecture. These models have been applied to tasks such as clinical question answering, entity extraction, and medical report summarization.
Despite these advances, the use of LLMs for radiology report generation—especially with complex imaging modalities such as CT scans—remains limited and largely underexplored.

\subsection{Radiology Report Generation}
Radiology report generation has been extensively studied in the context of chest X-ray imaging. Two widely used datasets in this domain are MIMIC-CXR~\cite{johnson2019mimic} and IU X-Ray~\cite{demner2016preparing}, both of which provide paired 2D chest X-ray images (typically including posteroanterior and lateral views) and corresponding free-text reports. However, these datasets lack detailed annotations of clinical findings, which limits their utility for structured or explainable report generation.
Hamamci et al.~\cite{hamamci2024developing} introduced a large-scale 3D chest CT dataset that supports multiple tasks, including relevant case retrieval, multi-abnormality detection, and radiology report generation. In a different domain, Lei et al.~\cite{lei2024autorg} released a brain MRI dataset containing both segmentation masks of abnormal regions and manually written reports, enabling spatially grounded report generation.

\section{Material and Method}
\subsection{Participants}
\label{subsec:Participants}

The data used in this study were obtained from the UF Health Integrated Data Repository (IDR)\footnote{\url{https://idr.ufhealth.org/}}. This study was approved by the Institutional Review Board (IRB202100401), and the requirement for written informed consent was waived. All procedures were conducted in accordance with the Declaration of Helsinki.

\begin{figure}[htbp]
    \centering
    \includegraphics[width=\textwidth]{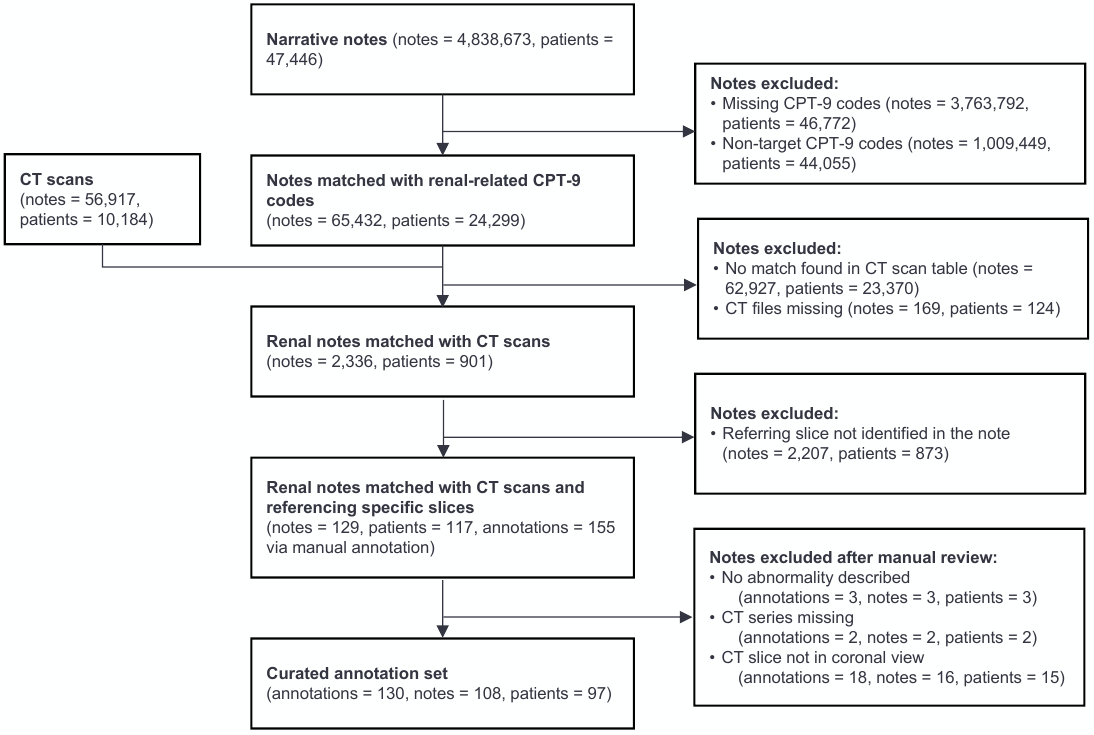}
    \caption{
    Workflow for filtering renal CT reports and scans. Reports were filtered based on renal-related CPT-9 codes, scan availability, and explicit references to individual CT slices. Reports referencing non-coronal slices or lacking abnormalities were excluded. Some reports contained multiple slice references, resulting in 130 manual annotations from 97 patients.
    }
    \label{fig:renal_annotation_flowchart}
\end{figure}

Figure~\ref{fig:renal_annotation_flowchart} illustrates the full pipeline for filtering renal CT reports and scans. Participants were identified through a multi-stage filtering process based on narrative radiology reports and their corresponding CT scans. Initially, all available renal CT reports between December 2, 2011 and August 24, 2024 were extracted and filtered using renal-related CPT-9 codes (74160, 74170, 74175, 74177, 74178), which correspond to abdominal and pelvic CT scans performed with or without intravenous contrast. Reports matching these codes were considered indicative of renal imaging.
These reports were then linked to the corresponding CT scan records. Pairs that could not be matched to an existing CT study or lacked available scan data were excluded. From the remaining report–scan pairs, we selected those explicitly referencing a specific CT slice.

Some reports referenced multiple distinct image slices, each associated with a different abnormality; thus, a single report could yield multiple annotations. These slice-referenced findings were manually annotated, resulting in 155 annotations derived from 129 reports. After excluding entries with no reported abnormalities, missing CT series, or non-coronal slices, we obtained a curated set of 130 annotations from 108 reports across 97 patients.

Table~\ref{table:demographic_summary} summarizes the patient demographics and lesion characteristics of the final cohort. Note that diagnosis labels differ between the patient level and lesion level. Patient-level diagnoses were extracted from all available ICD codes in each patient's medical history, while lesion-level annotations are restricted to abnormalities explicitly referenced by individual CT slices. Consequently, some diagnoses may appear at the patient level but be absent from the lesion-level summary.

\begin{table}[htbp]
    \centering
    \footnotesize
    \caption{Summary of patient and lesion characteristics.}
    \label{table:demographic_summary}

    \begin{minipage}[t]{0.45\linewidth}
    \footnotesize
    \centering
    \textbf{Patient characteristics (n = 97)} \\
    \vspace{0.3em}
    \begin{tabular}{ll}
        \toprule
        \textbf{Attribute} & \textbf{Value} \\
        \midrule
        \textbf{Age (years)} & 71.56 ± 6.34 \\
        \midrule
        \textbf{Race} & \\
        \quad Black & 30 \\
        \quad White & 66 \\
        \quad Other & 1 \\
        \midrule
        \textbf{Ethnicity} & \\
        \quad Not Hispanic & 96 \\
        \quad Unknown & 1 \\
        \midrule
        \textbf{Gender} & \\
        \quad Female & 51 \\
        \quad Male & 46 \\
        \midrule
        \textbf{Nephrectomy} & \\
        \quad Performed & 7 \\
        \quad Not Performed & 90 \\
        \midrule
        \textbf{Diagnosis Labels} & \\
        \quad Cancer & 16 \\
        \quad Tumor & 11 \\
        \quad Cyst & 88 \\
        \quad Other & 79 \\
        \midrule
        \textbf{CT Scan Counts} & \\
        \quad One scan & 69 \\
        \quad Two scans & 23 \\
        \quad Three scans & 5 \\
        \bottomrule
    \end{tabular}
    \end{minipage}
    \hfill
    \begin{minipage}[t]{0.52\linewidth}
    \centering
    \footnotesize
    \textbf{Lesion characteristics (n = 130)} \\
    \vspace{0.3em}
    \begin{tabular}{ll}
        \toprule
        \textbf{Attribute} & \textbf{Value} \\
        \midrule
        \textbf{Location} & \\
        \quad Right kidney & 68 \\
        \quad Left kidney & 61 \\
        \quad Unknown & 1 \\
        \midrule
        \textbf{Lesion Size (cm)} & 1.71 ± 1.20 \\
        \quad Known cases & 112 \\
        \quad Unknown & 18 \\
        \midrule
        \textbf{Growth Pattern} & \\
        \quad Exophytic & 26 \\
        \quad Endophytic & 2 \\
        \quad Unknown & 102 \\
        \midrule
        \textbf{Attenuation} & \\
        \quad Hypoattenuating & 43 \\
        \quad Hyperattenuating & 30 \\
        \quad Isoattenuating & 6 \\
        \quad Unknown & 51 \\
        \midrule
        \textbf{Enhancement} & \\
        \quad Enhancement & 26 \\
        \quad Non-enhancement & 10 \\
        \quad Unknown & 94 \\
        \midrule
        \textbf{Cyst} & \\
        \quad Present & 78 \\
        \quad Absent & 52 \\
        \midrule
        \textbf{Mass} & \\
        \quad Present & 15 \\
        \quad Absent & 115 \\
        \midrule
        \textbf{Tumor} & \\
        \quad Present & 7 \\
        \quad Absent & 123 \\
        \bottomrule
    \end{tabular}
    \end{minipage}
\end{table}

\subsection{Data Preprocessing}
\label{subsec:DataPreprocessing}
Our dataset comprises radiology reports, structured renal feature labels, and corresponding CT image slices. The data preprocessing steps are summarized in Figure~\ref{fig:model_pipeline}(a) and (b).

\subsubsection{Report}
The original radiology reports include sections such as clinical history, examination details, and imaging findings across multiple organ systems. For this study, we focused specifically on renal-related findings derived from CT observations. Given that our analysis centers on 2D CT slices, we extracted only sentences explicitly referencing specific image locations (e.g., “series 4 image 167”).  
To identify these slice-referenced sentences, we employed the Qwen 2.5 large language model~\cite{choi2023developing} with a customized prompt, followed by manual review to ensure precision and completeness, as illustrated in Figure~\ref{fig:model_pipeline}(a).
Sentences without explicit image references were excluded, even if they contained potentially relevant renal content.
Additionally, slice-referenced sentences describing findings unrelated to renal cancer progression (e.g., nephrolithiasis or hydronephrosis) were removed. This filtering ensured that each retained sentence aligned with a specific CT slice and remained within the clinical scope of renal cancer. The full prompt used for sentence extraction is provided in ~\ref{appendix:prompts}.

\subsubsection{Feature Labels}
Guided by physicians with sub-speciality expertise in the evaluation of renal tumors and informed by reference materials such as the Radiology Assistant~\cite{reinhard2016solid} and RadReport~\cite{smith2021stan}, we defined eight renal feature labels: position, size, growth pattern (exophytic/endophytic), attenuation, enhancement, cyst, mass, and tumor. Following the annotation protocol described in~\cite{choi2023developing}, a large language model was used to extract and structure these labels from relevant sentences, followed by manual verification.
When lesion type (e.g., cyst, mass, or tumor) could not be directly inferred, physician experts adjudicated the findings on a case-by-case basis. Feature values not explicitly mentioned were marked as \texttt{unknown}. For qualitative size descriptions (e.g., “subcentimeter”), values were standardized to 0.5 cm for consistency.

Several features—especially growth pattern and enhancement—were frequently omitted in original reports, leading to a high prevalence of \texttt{Unknown} entries (see Table~\ref{table:demographic_summary}). We intentionally chose not to impute or assign default values. Preserving missingness reflects the ambiguity inherent in real-world reports and allows downstream models to be trained and evaluated under realistic conditions.
The full prompt used for feature extraction is provided in Appendix~\ref{appendix:prompts}.

\subsubsection{CT Slice}
2D slices were extracted from the 3D CT volumes based on explicit references in the radiology reports (e.g., “series 4 image 167 ”). Slice positions were determined by sorting the DICOM series using the \texttt{SliceLocation} attribute, followed by index-based matching to the referenced image number. Each slice was aligned with its corresponding sentence and feature labels to form complete data triplets: (report sentence, feature label, CT slice).
We also standardized the spatial resolution of all CT slices by resizing each to a fixed dimension of $512 \times 512$ pixels. For slices larger than this size, we applied center cropping; for smaller ones, we padded the surrounding region with zeros. This resizing strategy ensured consistency across inputs while preserving lesion scale and spatial integrity—particularly important for size-sensitive tasks. 
In addition to spatial normalization, we applied standard intensity preprocessing. Raw CT scans were stored in Hounsfield Units (HU), spanning a wide dynamic range that includes soft tissue, bone, and air. To suppress irrelevant structures and emphasize renal parenchyma and lesions, we clipped HU values to a fixed window (window width = 400, window level = 50), as is standard in abdominal imaging. 
All CT slices were manually reviewed to confirm alignment with the renal region. Slices that were incorrectly referenced or anatomically misaligned in the original report were excluded.

\subsection{Data Splitting and Label Balance}
\label{subsec:Dataset}
A 5-fold cross-validation strategy was adopted to ensure robustness and generalizability. Data splitting was performed at the annotation level, where each sample corresponds to a unique triplet of (report sentence, label, CT slice). This approach enables stratification across a larger number of samples while preserving label diversity within each fold.
To mitigate sampling bias, we applied stratified sampling~\cite{sechidis2011stratification, szymanski2017network} based on renal feature distributions. For highly imbalanced features—such as the “Exophytic” category, which included only one “endophytic” case—we manually ensured that the minority class was represented in both training and validation sets.
The detailed label distribution for a representative fold is provided in Appendix~\ref{appendix:fold_distribution}.

\subsection{Model}
\label{subsec:model}
To improve clinical interpretability, we employ a two-stage modeling pipeline (Figure~\ref{fig:model_pipeline}). In the first stage, clinical features are detected from 2D CT slices using dedicated computer vision models. In the second stage, the detected features—together with the corresponding CT slice—are input into a fine-tuned VLM to generate a report. This modular architecture supports both structured feature evaluation and flexible, context-aware report generation.

\subsubsection{Feature Detection Model}
We use a ResNet backbone~\cite{he2016deep} as the base encoder for CT slice-level clinical feature detection. Each renal feature label is treated as an independent task:
\begin{itemize}[itemsep=0pt, parsep=0pt]
\item \textit{Binary classification}: Position, Enhancement, Cyst, Mass
\item \textit{Multi-class classification}: Attenuation (hypo-, hyper-, iso-attenuating)
\item \textit{Regression}: Size (cm)
\item \textit{Anomaly detection}: Tumor, Exophytic
\end{itemize}
The \textit{Exophytic} and \textit{Tumor} categories are highly imbalanced in our dataset (e.g., only one \textit{endophytic} case among 130 lesions), making conventional supervised classification ineffective. This imbalance may partly arise from reporting conventions, as radiologists at our institution rarely use the term “endophytic” explicitly.  For these features, we employ anomaly detection techniques~\cite{pang2021deep} trained exclusively on negative examples.

\subsubsection{Report Generation Model}
We used Qwen-VL 2.5~\cite{bai2023qwen} and LLaMA-3.2-Vision~\cite{dubey2024llama} as the backbone VLMs for radiology report generation. To assess the contribution of different input modalities, we evaluated three configurations: (1) structured clinical features only, (2) 2D CT slice images only, and (3) both modalities combined. Each model was tested in both fine-tuned (ft) and zero-shot (zs) settings, resulting in variants such as \texttt{ft-image}, \texttt{ft-feature}, and \texttt{ft-both}. Fine-tuning was conducted using LLAMA-Factory~\cite{zheng2024llamafactory} with instruction-style supervision, where the input consists of a prompt (containing clinical features and/or an image) and the output is the corresponding expert-written report sentence.
All experiments followed a consistent training and evaluation protocol. Specific prompt formats for each modality setting are detailed in Appendix~\ref{appendix:prompts}.

\subsection{Data Augmentation}
\label{subsec:Data Augmentation}
To improve model robustness and leverage spatial continuity in CT volumes, we applied augmentation strategies inspired by prior work~\cite{alzu2022kidney}. Specifically, for each annotated 2D CT slice, we included its adjacent slices at offsets of $-1$ and $+1$, effectively tripling the training data and introducing slight anatomical variation without altering underlying pathology.  For the feature detection model, we also experimented with different window level settings to enhance contrast and highlight renal abnormalities. After intensity clipping, voxel values were rescaled to the range $[-1, 1]$ using min-max normalization to promote training stability and improve convergence.

\subsection{Metrics}
\label{subsec:Metrics}
\subsubsection{Clinical Feature Evaluation}
To evaluate the performance of the feature extraction model, we adopt standard metrics tailored to each task type. For classification tasks—such as detecting cysts, masses, or tumors—we report Accuracy, Precision, Recall, and F1-score. To account for class imbalance (e.g., the tumor class has very few positive samples), we additionally report the Area Under the Receiver Operating Characteristic Curve (AUC-ROC), which offers a more reliable measure of discriminatory performance in imbalanced scenarios.
For regression tasks, such as predicting lesion size, we use Mean Squared Error (MSE) to quantify the average deviation between predicted and ground truth sizes (in cm). This metric captures both systematic and random errors in continuous predictions.

\subsubsection{Report Generation Evaluation}
To evaluate the similarity of generated radiology report sentences, we used three standard NLG metrics: BLEU~\cite{papineni2002bleu}, ROUGE-L~\cite{lin2004rouge}, and METEOR~\cite{banerjee2005meteor}. These metrics assess lexical similarity between generated and reference sentences by capturing n-gram overlap, longest common subsequence, and semantic alignment with synonym matching, respectively. In practice, we observe that different implementations of these metrics—such as those provided by the COCO Caption toolkit~\cite{chen2015microsoft}, NLTK~\cite{bird2009natural}, and Hugging Face—can yield slightly different numerical values due to variations in preprocessing, tokenization, and smoothing strategies. To ensure consistency and reproducibility, we adopt the implementations provided by the Hugging Face \texttt{evaluate} library~\cite{lhoest2021datasets} in all experiments.

\section{Results}
\label{sec:Results}

\subsection{Implementation Details}
\label{subsec:ExperimentalSetup}

For clinical feature detection, we performed a grid search across multiple ResNet variants (ResNet-18, ResNet-50, and ResNet-101), each initialized with ImageNet-pretrained weights. Input images were preprocessed using the augmentation strategies described in Section~\ref{subsec:Data Augmentation}, including adjacent slice incorporation, intensity windowing, normalization, and resizing to a fixed resolution of $512 \times 512$ pixels.
Models were trained with a batch size of 16, a learning rate of 1e-4, and the Adam optimizer. The best checkpoint was selected based on the highest validation F1 score.

For report generation, we fine-tuned the Qwen2.5-VL and LLaMA3.2-Vision models using Low-Rank Adaptation (LoRA) with rank = 8. Structured clinical features were embedded into the input prompt alongside the CT slice image (see Appendix~\ref{appendix:prompts}). Training was performed with a learning rate of 1e-4, batch size of 4 (gradient accumulation = 8), and a cosine learning rate scheduler over 20 epochs. Mixed-precision training (bf16) was enabled, and evaluation was conducted every 20 steps. The optimal checkpoint was selected based on the highest ROUGE-L score on the validation set.

All experiments were conducted on a single NVIDIA H100 B200 GPU with 192 GB of memory. The implementation was based on PyTorch 2.2 and Hugging Face Transformers, with all random seeds fixed for reproducibility.

\subsection{Clinical Feature Detection Results}
\label{subsec:result_abnormality}

Table~\ref{table:metric_grouped} summarizes the performance of our clinical feature detection model compared to a random baseline across all annotated abnormality types.
The proposed model consistently outperformed the random baseline across all detection tasks. For most binary features,the model achieved gains in both AUC (e.g., 0.7063 vs. 0.3168 for \textit{Tumor}) and F1 score (0.7400 vs. 0.3510). Performance was particularly high for \textit{Enhancement} (F1 = 0.8859) and \textit{Tumor} (F1 = 0.7400). 
For the multi-class classification task of \textit{Attenuation}, the model achieved moderate results, with a precision of 0.6300 and an F1 score of 0.6167. 
Although the evaluation metrics for the \textit{Exophytic} label appear perfect, they are misleading. The dataset contains only one instance labeled as “endophytic”, which appears in both the training and validation sets. 
For lesion size estimation, the model reported a mean squared error (MSE) of 0.9959 cm. Further analysis revealed that the error was primarily driven by a single cross-validation fold containing several unusually large lesions that were underrepresented in the training set. In other folds, the model exhibited strong consistency and accuracy.

\begin{table}[htbp]
\centering
\footnotesize
\caption{Performance comparison between the proposed clinical feature detection model and a random baseline across all annotated features.}
\label{table:metric_grouped}
\begin{tabular}{llccccc}
\toprule
\textbf{Feature} & \textbf{Model} & \textbf{AUC} & \textbf{Accuracy} & \textbf{Precision} & \textbf{Recall} & \textbf{F1 / MSE} \\
\midrule
\multirow{2}{*}{Position}
& Random   & 0.4643 & 0.4618 & 0.4539 & 0.4603 & 0.4471 \\
& Proposed & \textbf{0.6669} & \textbf{0.7175} & \textbf{0.733} & \textbf{0.7262} & \textbf{0.7132} \\
\midrule
\multirow{2}{*}{Exophytic}
& Random   & 0.5043 & 0.4533 & 0.4217 & 0.3471 & 0.3500 \\
& Proposed & \textbf{1.0000} & \textbf{1.0000} & \textbf{1.0000} & \textbf{1.0000} & \textbf{1.0000} \\
\midrule
\multirow{2}{*}{Attenuation}
& Random   & --     & 0.3202 & 0.3304 & 0.3165 & 0.2929 \\
& Proposed & --     & \textbf{0.6827} & \textbf{0.6300} & \textbf{0.6385} & \textbf{0.6167} \\
\midrule
\multirow{2}{*}{Enhancement}
& Random   & 0.3175 & 0.4528 & 0.4533 & 0.3395 & 0.3552 \\
& Proposed & \textbf{0.8500} & \textbf{0.9083} & \textbf{0.9258} & \textbf{0.8717} & \textbf{0.8859} \\
\midrule
\multirow{2}{*}{Cyst}
& Random   & 0.5784 & 0.5562 & 0.5584 & 0.5633 & 0.5397 \\
& Proposed & \textbf{0.7615} & \textbf{0.7807} & \textbf{0.7879} & \textbf{0.7664} & \textbf{0.7607} \\
\midrule
\multirow{2}{*}{Mass}
& Random   & 0.3870 & 0.4529 & 0.4613 & 0.4044 & 0.3569 \\
& Proposed & \textbf{0.5199} & \textbf{0.9084} & \textbf{0.8062} & \textbf{0.6325} & \textbf{0.6649} \\
\midrule
\multirow{2}{*}{Tumor}
& Random   & 0.3168 & 0.4829 & 0.4916 & 0.3929 & 0.3510 \\
& Proposed & \textbf{0.7063} & \textbf{0.9620} & \textbf{0.7344} & \textbf{0.7460} & \textbf{0.7400} \\
\midrule
\multirow{2}{*}{Size
}
& Random   & -- & -- & -- & -- & -- \\
& Proposed & -- & -- & -- & -- & \textbf{0.9595} \\
\bottomrule
\end{tabular}
\end{table}

\subsection{Radiology Report Generation Results}

We evaluated the quality of generated radiology reports using both NLG metrics and clinical feature accuracy. NLG metrics compare generated reports against radiologist-written reference sentences by measuring lexical similarity (Table~\ref{table:nlp_metrics}). To evaluate clinical correctness, we automatically extracted structured features from generated reports using an LLM-based parser and compared them with ground truth labels. Because the generated reports were produced by an LLM, they exhibit consistent formatting, reduced linguistic variability, and structured presentation of clinical findings, which facilitated information extraction with an accuracy exceeding 98\%. This evaluation captures whether the generated text faithfully conveys key clinical findings (Table~\ref{tab:all_model_metrics_part1}). All results are reported as averages over five cross-validation.

\subsubsection{Textual Evaluation of Generated Report}
For fine-tuned Qwen models, NLG scores remained relatively consistent across different input settings (clinical features only, image only, or both).
In the zero-shot setting, models tend to produce full-length radiology reports (see Figure~\ref{fig:case_study}), which differ in structure and verbosity from concise, expert-authored references, resulting in lower NLG scores.
We also observed that smaller models, such as LLaMA-11B, failed to converge under our experimental settings. Effective adaptation required scaling to the 90B-parameter variant.

\subsubsection{Generated Clinical Feature Evaluation}
\label{subsubsec:generated_feature_evaluation}
Since clinical features were extracted directly from the \emph{generated reports}, no confidence probabilities were available, making AUC and other thresold-based metrics inapplicable.
From Table~\ref{tab:all_model_metrics_part1}, we observed that models relying solely on image input (e.g., \texttt{ft-image}, \texttt{zs-image} variants) showed reduced performance for several features, including \texttt{Position}, \texttt{Exophytic}, \texttt{Attenuation}, and \texttt{Enhancement}.
Interestingly, fine-tuned models often yielded lower clinical feature accuracy than their zero-shot counterparts.
Fine-tuned models exhibited more frequent hallucinations for features such as \texttt{Cyst}, \texttt{Mass}, and \texttt{Tumor}, where as zero-shot models were comparatively less affected.
Regarding the \texttt{Size\_cm} feature, evaluated using MSE, we observe no consistent correlation between model scale and performance. Even larger models like Qwen-32B or LLaMA3-90B failed to outperform smaller ones. In some cases, models generated implausible multi-dimensional size estimates despite receiving only 2D inputs.

\subsection{Case Study}
To better understand model behavior, we examined two representative cases (Figure~\ref{fig:case_study}). Overall, fine-tuned models tend to generate reports that are structurally closer to those written by radiologists. In contrast, the zero-shot model (\texttt{Qwen-32B\textsubscript{zs-both}}) produced lengthy, well-structured reports that resemble radiology templates. This observation aligns with the trends in NLG evaluation metrics.
Interestingly, the zero-shot model closely adhered to the input prompt, reproducing the provided information faithfully without extrapolating beyond it. On the other hand, fine-tuned models occasionally introduced new content not explicitly mentioned in the prompt. Some of this added content is correct, while some are factually inaccurate. For example, in Case 2, \texttt{Qwen-32B\textsubscript{ft-both}} correctly inferred that the lesion is located in the interpolar region, which matches the ground truth. However, it also hallucinates a three-dimensional lesion size, despite only the largest diameter being provided in the prompt. 
Moreover, we observed that the overall quality and factual correctness of generated reports heavily depended on the accuracy of the input prompt.
When models relied solely on the image (e.g., \texttt{Qwen-32B\textsubscript{ft-image}}), the generated reports often deviated more significantly from the ground truth.

\begin{table}[htbp]
\centering
\footnotesize
\begin{threeparttable}
\caption{Comparison of lexical performance in radiology report generation across model configurations and input modalities. “Qwen” refers to the Qwen2.5-VL series; “LLaMA” refers to the LLaMA-3.2-Vision series.}
\label{table:nlp_metrics}
\begin{tabular}{lp{1.5cm}p{1.5cm}cccc}
\toprule
\textbf{Model} & \textbf{Clinical Feature} & \textbf{2D Slice Image} & \textbf{BLEU-1} & \textbf{BLEU-4} & \textbf{ROUGE-L} & \textbf{METEOR} \\
\midrule
Qwen-32B\textsubscript{ft-both}    & \cmark & \cmark & 0.299 & 0.079 & 0.281 & 0.325 \\
Qwen-32B\textsubscript{ft-image}   & \xmark & \cmark & 0.293 & 0.082 & 0.253 & 0.323 \\
Qwen-32B\textsubscript{ft-feature} & \cmark & \xmark & 0.278 & 0.069 & 0.273 & 0.309 \\

Qwen-32B\textsubscript{zs-both}    & \xmark & \xmark & 0.020 & 0.003 & 0.091 & 0.060 \\
Qwen-32B\textsubscript{zs-feature}     & \cmark & \xmark & 0.018 & 0.003 & 0.043 & 0.089 \\
Qwen-32B\textsubscript{zs-image}     & \xmark & \cmark & 0.014 & 0.002 & 0.033 & 0.075 \\
\midrule
Qwen-7B\textsubscript{ft-both}     & \cmark & \cmark & 0.271 & 0.069 & 0.275 & 0.307 \\
Qwen-7B\textsubscript{ft-feature}  & \cmark & \xmark & 0.294 & 0.087 & 0.280 & 0.309 \\
Qwen-7B\textsubscript{ft-image}    & \xmark & \cmark & 0.306 & 0.079 & 0.275 & 0.330 \\
Qwen-7B\textsubscript{zs-both}     & \cmark & \cmark & 0.031 & 0.004 & 0.128 & 0.078 \\
\midrule
LLaMA-11B\textsubscript{ft-both}   & \cmark & \cmark & 0.001 & 0.000 & 0.017 & 0.101 \\
LLaMA-90B\textsubscript{ft-both}   & \cmark & \cmark & 0.269 & 0.054 & 0.253 & 0.286 \\
\bottomrule
\end{tabular}
\footnotesize
Note: Model names use the prefix ``ft-'' for fine-tuned and ``zs-'' for zero-shot. The suffixes ``image'', ``feature'', and ``both'' indicate which modality/modalities are included in the input prompt.
\end{threeparttable}
\end{table}

\begin{table}[htbp]
\centering
\footnotesize
\caption{Evaluation of clinical feature performance extracted from generated reports.}
\label{tab:all_model_metrics_part1}
\begin{tabular}{llrrrr}
\toprule
Model & Feature & Accuracy & Precision & Recall & F1 / MSE \\
\midrule
\multirow{8}{*}{Qwen-32B\textsubscript{ft-both}} 
& Position     & 0.7175 & 0.7330 & 0.7262 & 0.7132 \\
& Exophytic    & 0.6133 & 0.6000 & 0.4471 & 0.5033 \\
& Attenuation  & 0.4157 & 0.1600 & 0.2735 & 0.1998 \\
& Enhancement  & 0.5772 & 0.5583 & 0.3864 & 0.4503 \\
& Cyst         & 0.5687 & 0.5982 & 0.5974 & 0.5608 \\
& Mass         & 0.4541 & 0.5463 & 0.5690 & 0.3939 \\
& Tumor        & 0.9392 & 0.5767 & 0.5919 & 0.5841 \\
& Size\_cm     & --     & --     & --     & 1.7444 \\
\midrule
\multirow{8}{*}{Qwen-32B\textsubscript{ft-feature}} 
& Position     & 0.6938 & 0.7268 & 0.7041 & 0.6973 \\
& Exophytic    & 0.7367 & 0.7000 & 0.6007 & 0.6419 \\
& Attenuation  & 0.4794 & 0.2792 & 0.3098 & 0.2543 \\
& Enhancement  & 0.5594 & 0.5033 & 0.4114 & 0.4441 \\
& Cyst         & 0.4452 & 0.5371 & 0.5144 & 0.4229 \\
& Mass         & 0.3207 & 0.5324 & 0.5269 & 0.3066 \\
& Tumor        & 0.8333 & 0.4696 & 0.4400 & 0.4533 \\
& Size\_cm     & --     & --     & --     & 1.1081 \\
\midrule
\multirow{8}{*}{Qwen-32B\textsubscript{ft-image}} 
& Position     & 0.4487 & 0.4494 & 0.4420 & 0.4267 \\
& Exophytic    & 0.1833 & 0.4000 & 0.1086 & 0.1683 \\
& Attenuation  & 0.2946 & 0.3176 & 0.2061 & 0.2448 \\
& Enhancement  & 0.1244 & 0.4000 & 0.1043 & 0.1583 \\
& Cyst         & 0.4909 & 0.5377 & 0.5327 & 0.4853 \\
& Mass         & 0.3735 & 0.5094 & 0.4981 & 0.3436 \\
& Tumor        & 0.9310 & 0.4728 & 0.4917 & 0.4820 \\
& Size\_cm     & --     & --     & --     & 1.9741 \\
\midrule
\multirow{8}{*}{Qwen-32B\textsubscript{zs-both}} 
& Position     & 0.7175 & 0.7330 & 0.7262 & 0.7132 \\
& Exophytic    & 1.0000 & 1.0000 & 1.0000 & 1.0000 \\
& Attenuation  & 0.5136 & 0.1712 & 0.3333 & 0.2254 \\
& Enhancement  & 0.7989 & 0.6300 & 0.6281 & 0.6154 \\
& Cyst         & 0.7658 & 0.7771 & 0.7482 & 0.7403 \\
& Mass         & 0.8702 & 0.6548 & 0.6113 & 0.6038 \\
& Tumor        & 0.9620 & 0.7344 & 0.7460 & 0.7400 \\
& Size\_cm     & --     & --     & --     & 1.0987 \\
\midrule
\multirow{8}{*}{Qwen-32B\textsubscript{zs-feature}} 
& Position     & 0.7175 & 0.7330 & 0.7262 & 0.7132 \\
& Exophytic    & 1.0000 & 1.0000 & 1.0000 & 1.0000 \\
& Attenuation  & 0.5136 & 0.1712 & 0.3333 & 0.2254 \\
& Enhancement  & 0.8389 & 0.6449 & 0.6074 & 0.5993 \\
& Cyst         & 0.7735 & 0.7848 & 0.7573 & 0.7494 \\
& Mass         & 0.8862 & 0.7448 & 0.6350 & 0.6592 \\
& Tumor        & 0.9469 & 0.6137 & 0.6420 & 0.6259 \\
& Size\_cm     & --     & --     & --     & 1.0987 \\
\bottomrule
\end{tabular}
\end{table}

\begin{table}[htbp]
\ContinuedFloat
\centering
\footnotesize
\caption{Evaluation of clinical feature performance extracted from generated reports. (continued)}
\label{tab:all_model_metrics_part2}
\begin{tabular}{llrrrr}
\toprule
Model & Feature & Accuracy & Precision & Recall & F1 / MSE \\
\midrule
\multirow{8}{*}{Qwen-32B\textsubscript{zs-image}} 
& Position     & 0.5040 & 0.5207 & 0.4990 & 0.4962 \\
& Exophytic    & 0.1667 & 0.2667 & 0.1000 & 0.1405 \\
& Attenuation  & 0.4289 & 0.3090 & 0.2862 & 0.2919 \\
& Enhancement  & 0.2483 & 0.3275 & 0.2667 & 0.2486 \\
& Cyst         & 0.5147 & 0.4803 & 0.4918 & 0.4739 \\
& Mass         & 0.5097 & 0.4654 & 0.4083 & 0.3883 \\
& Tumor        & 0.8925 & 0.4719 & 0.4717 & 0.4709 \\
& Size\_cm     & --     & --     & --     & 3.1239 \\
\midrule
\multirow{8}{*}{Qwen-7B\textsubscript{ft-both}} 
& Position     & 0.7175 & 0.7330 & 0.7262 & 0.7132 \\
& Exophytic    & 0.6533 & 0.7800 & 0.6271 & 0.6833 \\
& Attenuation  & 0.4088 & 0.3041 & 0.2696 & 0.2361 \\
& Enhancement  & 0.5844 & 0.5875 & 0.4083 & 0.4596 \\
& Cyst         & 0.5445 & 0.6361 & 0.6064 & 0.5218 \\
& Mass         & 0.3340 & 0.5563 & 0.5772 & 0.3190 \\
& Tumor        & 0.9307 & 0.4726 & 0.4913 & 0.4817 \\
& Size\_cm     & --     & --     & --     & 1.2672 \\
\midrule
\multirow{8}{*}{Qwen-7B\textsubscript{ft-feature}} 
& Position     & 0.6921 & 0.7346 & 0.6980 & 0.7020 \\
& Exophytic    & 0.8683 & 0.7000 & 0.6857 & 0.6923 \\
& Attenuation  & 0.4580 & 0.3274 & 0.3042 & 0.2739 \\
& Enhancement  & 0.6467 & 0.5124 & 0.4357 & 0.4667 \\
& Cyst         & 0.5648 & 0.6045 & 0.6054 & 0.5575 \\
& Mass         & 0.4509 & 0.5704 & 0.6732 & 0.4142 \\
& Tumor        & 0.8929 & 0.4716 & 0.4720 & 0.4715 \\
& Size\_cm     & --     & --     & --     & 1.8435 \\
\midrule
\multirow{8}{*}{Qwen-7B\textsubscript{ft-image}} 
& Position     & 0.4504 & 0.4433 & 0.4403 & 0.4355 \\
& Exophytic    & 0.4467 & 0.5333 & 0.4329 & 0.4421 \\
& Attenuation  & 0.3488 & 0.3244 & 0.2482 & 0.2679 \\
& Enhancement  & 0.1022 & 0.3000 & 0.0650 & 0.1067 \\
& Cyst         & 0.4961 & 0.5294 & 0.5295 & 0.4937 \\
& Mass         & 0.3942 & 0.4845 & 0.4609 & 0.3404 \\
& Tumor        & 0.9170 & 0.4727 & 0.4846 & 0.4782 \\
& Size\_cm     & --     & --     & --     & 1.9127 \\
\midrule
\multirow{8}{*}{Qwen-7B\textsubscript{zs-both}} 
& Position     & 0.7175 & 0.7330 & 0.7262 & 0.7132 \\
& Exophytic    & 1.0000 & 1.0000 & 1.0000 & 1.0000 \\
& Attenuation  & 0.5136 & 0.1712 & 0.3333 & 0.2254 \\
& Enhancement  & 0.9083 & 0.7258 & 0.6717 & 0.6859 \\
& Cyst         & 0.7496 & 0.7656 & 0.7139 & 0.7182 \\
& Mass         & 0.8787 & 0.7090 & 0.6886 & 0.6846 \\
& Tumor        & 0.9395 & 0.6591 & 0.7340 & 0.6836 \\
& Size\_cm     & --     & --     & --     & 1.0987 \\
\bottomrule
\end{tabular}
\end{table}
\begin{table}[htbp]
\ContinuedFloat
\centering
\footnotesize
\caption{Evaluation of clinical feature performance extracted from generated reports. (continued)}
\label{tab:all_model_metrics_part3}
\begin{tabular}{llrrrr}
\toprule
Model & Feature & Accuracy & Precision & Recall & F1 / MSE \\
\midrule
\multirow{8}{*}{LLaMA3-11B\textsubscript{ft-both}}  \\
& Position     & 0.5508 & 0.6400 & 0.5325 & 0.5549 \\
& Exophytic    & 0.4517 & 0.3417 & 0.2743 & 0.2941 \\
& Attenuation  & 0.1982 & 0.1866 & 0.1567 & 0.1398 \\
& Enhancement  & 0.3356 & 0.4014 & 0.2414 & 0.2681 \\
& Cyst         & 0.4247 & 0.5090 & 0.5002 & 0.3632 \\
& Mass         & 0.2468 & 0.4442 & 0.4098 & 0.2307 \\
& Tumor        & 0.9458 & 0.4729 & 0.5000 & 0.4860 \\
& Size\_cm     & --     & --     & --     & 4.3634 \\
\midrule
\multirow{8}{*}{LLaMA3-90B\textsubscript{ft-both}} 
& Position     & 0.7175 & 0.7330 & 0.7262 & 0.7132 \\
& Exophytic    & 0.9333 & 0.8000 & 0.8000 & 0.8000 \\
& Attenuation  & 0.4686 & 0.1679 & 0.3019 & 0.2151 \\
& Enhancement  & 0.7783 & 0.6258 & 0.5650 & 0.5809 \\
& Cyst         & 0.5688 & 0.6092 & 0.6048 & 0.5412 \\
& Mass         & 0.3815 & 0.5613 & 0.6027 & 0.3501 \\
& Tumor        & 0.9229 & 0.5758 & 0.5830 & 0.5792 \\
& Size\_cm     & --     & --     & --     & 1.0864 \\
\bottomrule
\end{tabular}
\end{table}

\begin{figure}[htbp]
    \centering
    \includegraphics[width=\textwidth]{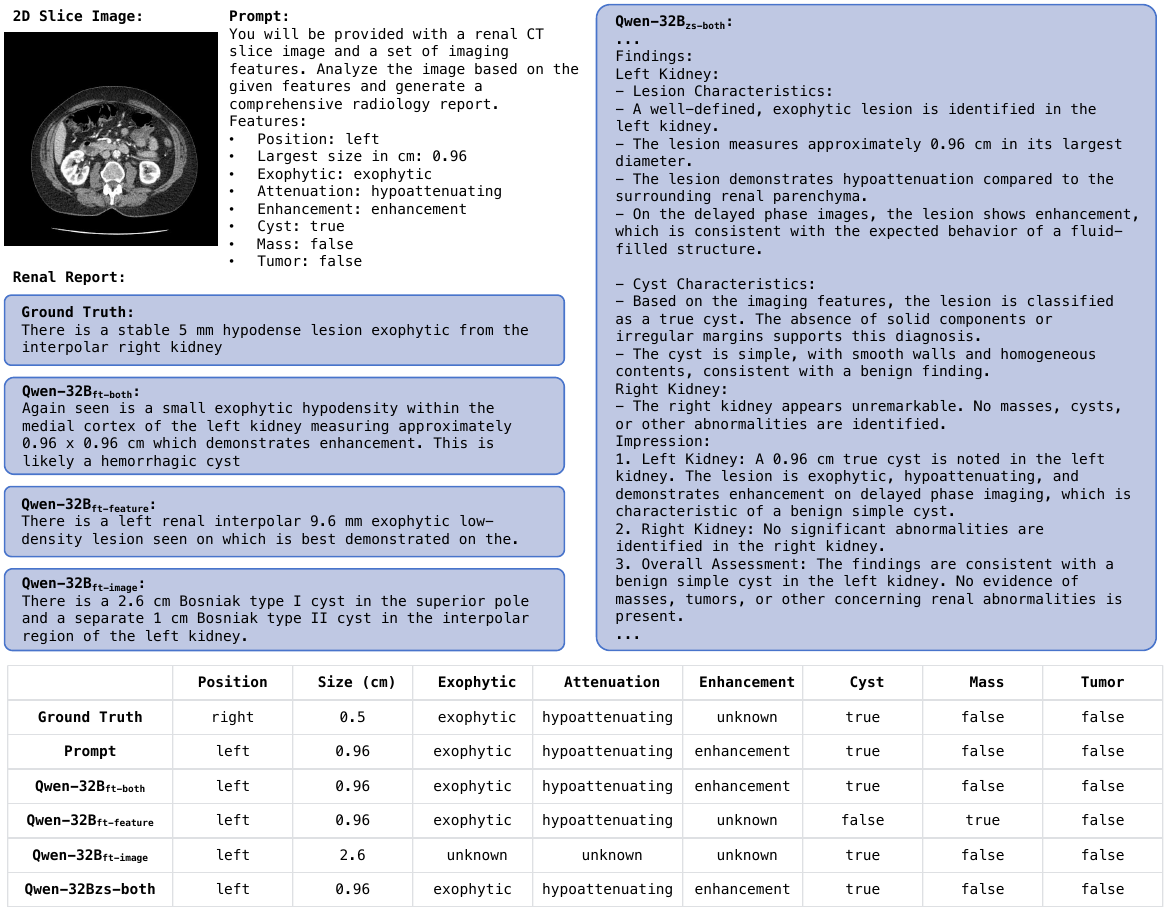}
    \caption{Comparison of model-generated radiology reports and extracted features under different input modalities (case 1).}
    \label{fig:case_study}
\end{figure}

\begin{figure}[htbp]
    \centering
    \ContinuedFloat
    \includegraphics[width=\textwidth]{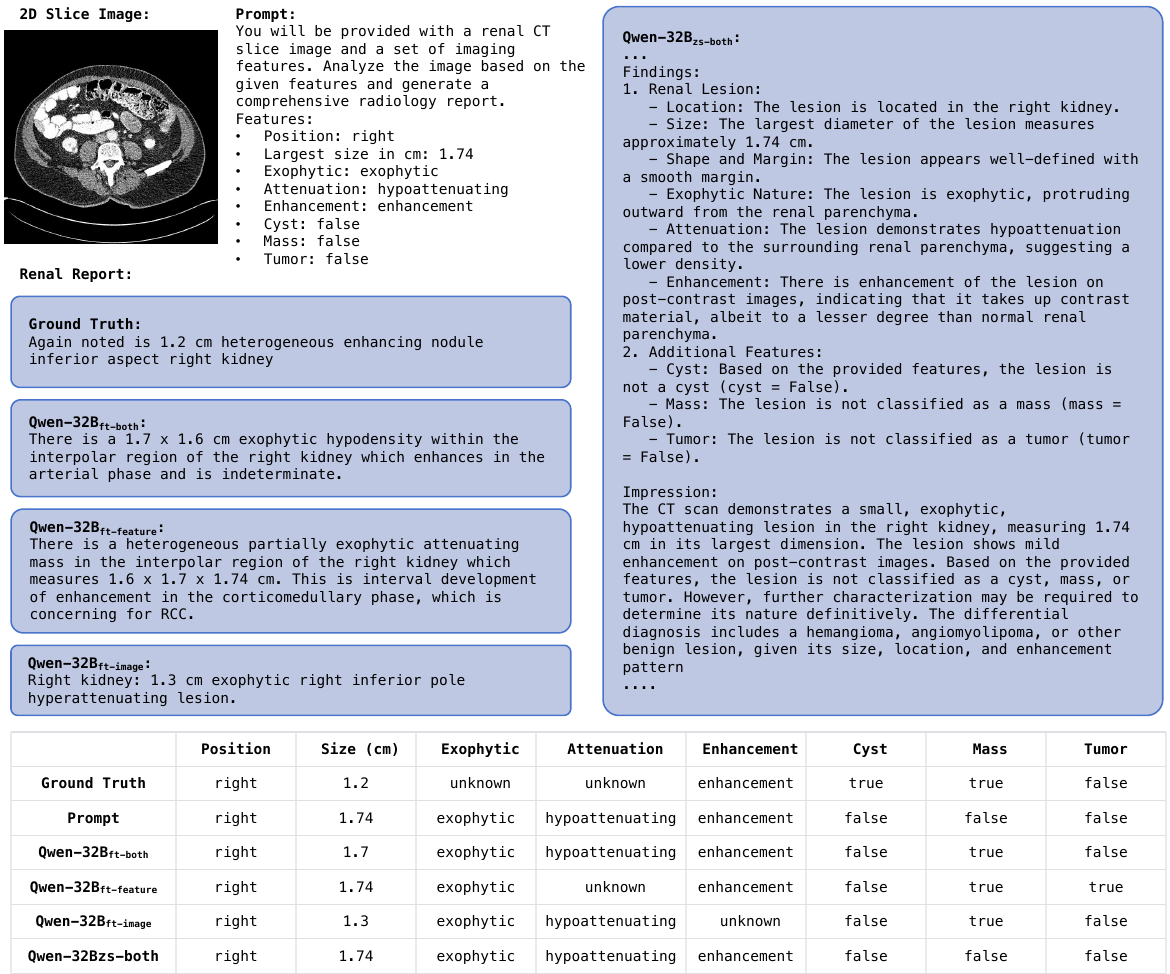}
    \caption{Comparison of model-generated radiology reports and corresponding extracted features across different input modalities (case 2).}
    \label{fig:case_study2}
\end{figure}

\section{Discussion}
\label{sec:Discussion}

\subsection{Clinical Feature Detection}
Overall, the proposed model consistently outperformed the random baseline across all feature detection tasks, demonstrating robust learning and strong generalization.
For most binary features—such as \textit{Position}, \textit{Cyst}, and \textit{Tumor}—the model achieved notable gains in both AUC  and F1 score, showing its effectiveness in identifying clinically relevant findings. Performance was particularly strong for the \textit{Enhancement} and \textit{Tumor} categories, both of which are critical for lesion characterization and cancer diagnosis.

By contrast, the moderate performance on \textit{Attenuation} highlights the difficulty of this task. This limitation is likely due to overlapping HU values among hypo-, iso-, and hyperattenuating tissues, combined with the scarcity of explicit attenuation descriptions in the source reports. Nevertheless, the model achieved more than a twofold improvement in accuracy compared to the baseline, suggesting that it effectively captures meaningful attenuation patterns despite data constraints.
The apparently perfect metrics for the \textit{Exophytic} category are misleading. With only one “endophytic” instance present across training and validation, this likely caused overfitting via memorization, meaning the reported metrics do not reflect true generalization for this feature.
For lesion size estimation, further analysis revealed that one fold showed substantially higher error than the others. This was primarily due to the presence of several unusually large lesions in the validation set, which were underrepresented in the corresponding training data. A more balanced distribution of lesion sizes in the training set could therefore improve regression performance.

\subsection{Radiology Report Generation}

The results highlight several important trends in report generation. First, models relying solely on image input struggled to capture detailed clinical findings. When models relied only on the image (e.g., \texttt{Qwen-32B\textsubscript{ft-image}}), the generated reports deviated more significantly from the ground truth. This suggests that our two-stage framework—first extracting structured clinical features and then generating the report—is more reliable than directly generating text from images, particularly in terms of clinical fidelity.  
Second, fine-tuned models often underperformed their zero-shot counterparts in clinical feature accuracy. 
We hypothesize that this may be due to fine-tuning on ground-truth reports, which encourages the model to imitate the linguistic style of radiologists. Such stylistic imitation can lead to the omission of explicit clinical details or the reproduction of incorrect information, thereby negatively affecting the expression of key clinical features.
Moreover, the hallucinating in fine-tuned models may arise from memorized report templates that promote overgeneralization and misrepresentation of clinical content. Zero-shot models were comparatively less affected by this problem.  
Lesion size estimation remains a particularly challenging task. In several cases, models generated 3-dimensional size despite receiving only inputs 2-dimensional size.
Finally, no consistent relationship was observed between model scale and clinical accuracy. This may reflect the limited size of our dataset, which restricts the benefits of scaling to larger models. 
Overall, these findings suggest that while large vision–language models are capable of producing fluent reports, challenges remain in ensuring factual correctness and clinical fidelity.

\subsection{Limitations}
\label{sec:Limitations}

\paragraph{Data Imbalance and Dataset Size}
Our dataset is relatively small and exhibits substantial class imbalance, particularly for rare feature labels such as \textit{endophytic} and \textit{tumor}. For example, the seemingly perfect performance on the \textit{Exophytic} category is due to a single minority instance appearing in both training and validation sets, resulting in inflated metrics. 
In addition, the rarity of the \textit{endophytic} label may reflect reporting conventions rather than true clinical frequency, as radiologists at our institution may not consistently use this descriptor. 
Future work should incorporate multi-center data annotated by diverse radiologists to better capture reporting heterogeneity and improve generalizability. 
Expanding the dataset and improving class balance will be essential for obtaining more reliable performance estimates~\cite{johnson2019mimic, irvin2019chexpert}.

\paragraph{Incomplete Annotations from Reports}
All feature labels were extracted from narrative radiology reports, which frequently omit attributes that are not considered clinically significant. As a result, certain features—such as \textit{enhancement} and \textit{attenuation}—contain a large proportion of “unknown” values. Although we intentionally preserved these missing entries to reflect real-world reporting practices, they introduce challenges in both model training and evaluation.

\paragraph{Non-uniform Slice Thickness}
In this work, we did not standardize the slice thickness of CT scans in physical space. 
This variability may lead to inconsistent spatial coverage during data augmentation, 
which can in turn affect lesion size estimation and the model’s spatial reasoning. 
Future work should incorporate slice spacing metadata from DICOM headers or resample all scans 
to a uniform voxel resolution to ensure consistent 3D context.

\paragraph{2D Slice Limitation}
Our current pipeline operates on individual 2D CT slices. While this design simplifies preprocessing and alignment with sentence-level annotations, it limits the model’s capacity to capture spatial continuity. Many renal lesions span multiple slices, and extending the framework to volumetric (3D) modeling could improve performance by leveraging richer spatial context.

\paragraph{Loss of Clinical Feature Fidelity in Generated Reports}
While the generated reports often appear fluent and well-structured, they may fail to accurately represent the intended clinical features, particularly in fine-tuned models. This discrepancy becomes evident during automatic clinical feature extraction, where key findings are occasionally omitted or hallucinated (e.g., incorrect lesion size, false tumor attribution). Such deviations reduce the clinical reliability of the generated reports and underscore a gap between surface-level textual quality and underlying clinical fidelity.

\section{Conclusion}
\label{sec:Conclusion}

This study presents a clinically informed, two-stage framework for renal radiology report generation from 2D CT slices. The first stage detects structured clinical features, and the second stage generates corresponding free-text reports conditioned on both the CT image and the detected features. This design enables the model to simultaneously produce structured and unstructured outputs, closely aligning with the dual needs of radiology workflows—supporting both quantitative feature analysis and narrative documentation.

While the current system is limited to 2D slices, it lays the groundwork for more comprehensive and clinically realistic solutions. In future work, we plan to extend the framework to full 3D CT volumes to capture richer anatomical context and enable more robust report generation. Additional efforts may focus on improving the clinical fidelity of generated reports through fact-checking modules, constrained decoding, or feature-aware training objectives. Expanding the dataset and improving label completeness—potentially through multi-source annotation or expert refinement—will also be critical for addressing data sparsity and class imbalance.

\section*{Funding}
Not available.

\section*{Author contributions}
RL: Conceptualization, Data curation, Methodology, Writing – original draft.  
ZF: Conceptualization, Data curation.  
JP: Methodology, Validation.  
CS: Methodology, Validation.  
RT: Formal analysis, Supervision, Writing – review \& editing.  
BS: Data curation.  
JX: Conceptualization, Supervision, Writing – review \& editing.

\section*{Competing interests}
The authors declare that they have no competing interests.

\section*{Data availability}
Clinical CT images and radiology reports were obtained from the UF Health Integrated Data Repository (IRB202400720). 
Due to HIPAA regulations and institutional data-use agreements, the raw clinical data cannot be shared publicly. 
To support reproducibility, we provide the full preprocessing and training code at \url{https://github.com/renjie-liang/two-stage-renal-ct-report-generation}, 
along with a small synthetic dataset that follows the same structure and labeling schema as the real data.

\section*{Declaration of generative AI and AI-assisted technologies in the writing process}
During the preparation of this manuscript, the authors used ChatGPT to improve the readability and language.  After using this tool/service, the author(s) reviewed and edited the content as needed and take(s) full responsibility for the content of the published article. No patient data or protected health information (PHI) were provided to the tool.

{

    \small
    \bibliography{main}
}

\newpage
\appendix

\section{Prompt Design}
\label{appendix:prompts}

We designed a series of prompts to guide large language models in three key components of our pipeline. First, sentence-level filtering was performed using LLM to identify renal-related content that references specific CT slices. Second, structured abnormality features were extracted from the filtered sentences, focusing on lesion characteristics relevant to renal cancer. Third, we fine-tuned the Qwen-VL 2.5 vision-language model for report generation, using paired inputs of CT slices and extracted features. All prompts are provided below in their original form.

\begin{tcolorbox}[colback=gray!5!white, colframe=gray!50!black, title=Prompt 1: Renal Sentence Extraction, fonttitle=\bfseries]
\footnotesize
\textbf{Task:}  
Identify and extract kidney/renal-related text snippets from a radiology CT report while excluding adrenal-related content.

\vspace{0.5em}
\textbf{1. Analyze the Report Structure:}  
Common sections include:
\begin{itemize}
    \item \texttt{HISTORY}
    \item \texttt{EXAM}
    \item \texttt{PRIOR STUDY}
    \item \texttt{FINDINGS}
    \item (Other possible sections)
\end{itemize}

\vspace{0.3em}
\textbf{2. Identify Renal-Relevant Terms:}
\begin{itemize}
    \item \textit{Include}: kidney, renal, nephro-, ureter, cyst, calculi, stone, hydronephrosis, parenchyma, cortex, medulla, atrophy, mass, tumor, lesion
    \item \textit{Exclude}: adrenal-related terms (e.g., adrenal, suprarenal)
\end{itemize}

\vspace{0.3em}
\textbf{3. Extract and Format Results:}
For each section, output:
\begin{itemize}
    \item Direct quotes of renal-relevant snippets, or
    \item \texttt{"none"} if no renal terms are found
\end{itemize}

\textbf{Output Template:}
\begin{lstlisting}[basicstyle=\ttfamily\footnotesize]
{
  "renal_extracts": {
    "HISTORY": extracted snippet or "none",
    "EXAM": extracted snippet or "none",
    "PRIOR STUDY": extracted snippet or "none",
    "FINDINGS": extracted snippet or "none"
  }
}
\end{lstlisting}
\end{tcolorbox}

\begin{tcolorbox}[colback=gray!5!white, colframe=gray!50!black, title=Prompt 2: Renal Feature Extraction, fonttitle=\bfseries]
\footnotesize

\textbf{Task:} Identify and extract kidney/renal-related abnormality information from a radiology CT report. Focus on lesions, masses, cysts, and tumors; exclude kidney stones and hydronephrosis.

\vspace{0.5em}
\textbf{Instructions:}
\begin{itemize}
    \item Detect all relevant abnormalities (lesion, mass, cyst, tumor).
    \item If no abnormality is found, return \texttt{{"Abnormality": false}}.
    \item Otherwise, return \texttt{{"Abnormality": true}} with extracted fields.
\end{itemize}

\vspace{0.3em}
\textbf{Fields to Extract:}
\begin{itemize}
    \item \textbf{Location:} Position (left/right)
    \item \textbf{Size:} Raw size string and standardized size in cm
    \item \textbf{Characteristics:} Exophytic, attenuation, enhancement
    \item \textbf{Classification:} Boolean flags for Lesion, Cyst, Mass, Tumor
    \item \textbf{Raw Fields:} Verbatim report text (prefixed with \texttt{Raw\_})
\end{itemize}

\textbf{Missing Values:} Use \texttt{"unknown"} for any unspecified field.

\textbf{Example Output Format:}
\begin{lstlisting}[basicstyle=\ttfamily\footnotesize]
{
 "Abnormality": true,
 "Abnormality_Info": [
   {
     "Position": "left",
     "Raw_Size": "3.2 * 2.8 cm",
     "Size_cm": 3.2,
     "Exophytic": "exophytic",
     "Attenuation": "Hyperattenuating",
     "Enhancement": "enhancement",
     "Lesion": true,
     "Cyst": true,
     "Mass": true,
     "Tumor": false,
   },
 ]
}
\end{lstlisting}

\textbf{Classification Rules:}  
Use domain-informed heuristics such as:
\begin{itemize}
    \item "denser than water" $\rightarrow$ Hyperattenuating
    \item "complex cystic mass" $\rightarrow$ Cyst: true, Mass: true
    \item Any mention suggestive of RCC (e.g., “suspicious for RCC”) $\rightarrow$ Tumor: true
\end{itemize}

\end{tcolorbox}

\begin{tcolorbox}[colback=gray!5!white, colframe=gray!50!black, title=Prompt 3: Report Generation, fonttitle=\bfseries]
\footnotesize

\textbf{System Prompt:}  
This system generates CT radiology reports specifically for the diagnosis of renal abnormalities (lesion, cyst, mass, tumor).

\vspace{0.5em}
\textbf{User Input:}  
You will be provided with a renal CT slice image and a set of imaging features. Analyze the image based on the given features and generate a comprehensive radiology report.

\textbf{Features:}
\begin{itemize}
    \item Position: \texttt{left}
    \item Largest size for the lesion (cm): \texttt{1.78}
    \item Exophytic: \texttt{exophytic}
    \item Attenuation: \texttt{hypoattenuating}
    \item Enhancement: \texttt{enhancement}
    \item Cyst: \texttt{true}
    \item Mass: \texttt{false}
    \item Tumor: \texttt{false}
\end{itemize}

\textbf{Image (CT scan slice):} \texttt{<image>}

\vspace{0.5em}
\textbf{Expected Output:}
\begin{quote}
Left kidney: There is a 3.6 x 3.4 x 2.9 cm exophytic, complex cystic and solid mass with focal punctate calcification identified at the superior pole of the left kidney. This mass extends into the left quadratus lumborum muscle with loss of fat planes of separation.
\end{quote}

\end{tcolorbox}

\section{Label Distribution}
\label{appendix:fold_distribution}

Table~\ref{table:fold_distribution} summarizes the distribution of renal feature labels in one representative fold of our 5-fold cross-validation setup. It reflects the class balance and prevalence of missing values that our stratified sampling strategy aimed to preserve.

\begin{table}[htbp]
    \centering
    \footnotesize
    \caption{Label Distribution in One Fold of the Cross-Validation Dataset}
    \label{table:fold_distribution}
    \begin{tabular}{lll}
        \toprule
        \textbf{Attribute} & \textbf{Training} & \textbf{Validation} \\
        \midrule
        \multicolumn{3}{l}{\textbf{Position}} \\
        Right kidney & 58 & 10 \\
        Left kidney & 46 & 15 \\
        Unknown & 1 & 0 \\
        \midrule
        \multicolumn{3}{l}{\textbf{Lesion Size (cm)}} \\
        Mean ± SD & 1.73 ± 1.26 & 1.61 ± 0.95 \\
        Max / Min / Median & 7.4 / 0.28 / 1.4 & 3.6 / 0.5 / 1.4 \\
        Unknown & 14 & 4 \\
        \midrule
        \multicolumn{3}{l}{\textbf{Growth Pattern}} \\
        Exophytic & 20 & 6 \\
        Endophytic & 1 & 1 \\
        Unknown & 84 & 18 \\
        \midrule
        \multicolumn{3}{l}{\textbf{Attenuation}} \\
        Hypoattenuating & 34 & 9 \\
        Hyperattenuating & 25 & 5 \\
        Isoattenuating & 4 & 2 \\
        Unknown & 42 & 9 \\
        \midrule
        \multicolumn{3}{l}{\textbf{Enhancement}} \\
        Enhancement & 22 & 4 \\
        Non-enhancement & 8 & 2 \\
        Unknown & 75 & 19 \\
        \midrule
        \multicolumn{3}{l}{\textbf{Cyst}} \\
        Present & 62 & 16 \\
        Absent & 43 & 9 \\
        \midrule
        \multicolumn{3}{l}{\textbf{Mass}} \\
        Present & 11 & 4 \\
        Absent & 94 & 21 \\
        \midrule
        \multicolumn{3}{l}{\textbf{Tumor}} \\
        Present & 6 & 1 \\
        Absent & 99 & 24 \\
        \bottomrule
    \end{tabular}
\end{table}

\end{document}